\documentclass[twocolumn]{IEEEtran}
\usepackage{amsmath}
\usepackage{amssymb}
\usepackage{float}
\usepackage{multirow}
\usepackage{graphicx}
\usepackage{color}
\usepackage{tikz} 

\newtheorem{thm}{Theorem}

\newtheorem{lem}{Lemma}
\newtheorem{prop}{Proposition}
\newtheorem{ex}{Example}

\floatstyle{ruled}
\newfloat{algorithm}{tbp}{loa}
\providecommand{\algorithmname}{Algorithm}
\floatname{algorithm}{\protect\algorithmname}
\usepackage{algorithm}
\usepackage{algorithmic}

\begin{document}

\title{Reinforcement Learning Approach to Estimation in Linear Systems}
\author{{\normalsize{Minyue Fu$^{1}$}} 
\thanks{$^1$School of Electrical Engineering and Computing, The University of Newcastle, University Drive, Callaghan, 2308, NSW, Australia.}
\thanks{E-mail: minyue.fu@newcastle.edu.au.}
}
\maketitle
\begin{abstract}
This paper addresses two important estimation problems for linear systems, namely system identification and model-free state estimation. Our focus is on ARMAX models with unknown parameters. We first provide a reinforcement learning algorithm for system identification with guaranteed consistency. This algorithm is then used to provide a novel solution to model-free state estimation. These results are then applied to solving the model-free LQG control problem in the reinforcement learning setting. 
\end{abstract}


\section{Introduction}
It is well known that {\em system identification} and {\em state estimation} are closely related learning problems for dynamic systems, with a rich history of research and rich set of methodologies; see, e.g., classical monographs \cite{Ljung,Soderstrom} for the former and \cite{Anderson,Goodwin} for the latter. The task of system identification is to estimate the system parameters, whereas that of state estimation is to provide an estimate of the state for a given system model. 

The main motivation for this paper is to understand {\em how to do state estimation without a system model}.  A simple approach is, of course, to estimate the system parameters first and then use them to estimate the state. But this approach is not suitable for on-line model-free state estimation where the estimates need to be updated recursively (or iteratively) along with the output measurement samples. That is, an online estimation algorithm is preferred. The second motivation for this paper is to know whether these estimation problems can be studied in the framework of reinforcement learning \cite{Silver,Bertsekas}. 

The system under study is the classical Auto-Regressive Moving-Average eXogenous (ARMAX) model with known orders but unknown parameters. We first consider the online system identification problem formulated in the reinforcement learning framework, and the objective is to provide a recursive (or iterative) estimate of the system parameters along with the update of the output measurement. By blending the tools of {\em instrumental variables} and {\em bootstrapping}, we provide a new recursive learning algorithm that globally optimises a cost function in the reinforcement learning setting and provides a convergent and consistent parameter estimate in the system identification setting at the same time. We then extend this algorithm to solve the model-free state estimation problem under a similar reinforcement learning setting and give an asymptotically optimal state estimate in the Kalman filtering sense. The reinforcement learning algorithms for system identification and state estimation will then be used to solve the classical linear quadratic Gaussian (LQG) control problem for an ARMAX model with unknown parameters. The solution is a reinforcement learning algorithm for model-free LQG control. 

The contributions of the paper are summarised below:
\begin{itemize}
\item Reformulation and reinterpretation of the classical system identification tools (least-squares, instrumental variables, bootstrapping...) in the framework of reinforcement learning;
\item New recursive parameter estimation algorithm for system identification with consistency;
\item Reinforcement learning algorithm for model-free state estimation;
\item Application to model-free linear quadratic Gaussian (LQG) control.
\end{itemize}


\section{Problem Statements}

\subsection{System Model}
In this paper, we consider a system with the following stationary ARMAX model \cite{Ljung}:
\begin{align}
&\  y_k + a_1y_{k-1}+\ldots + a_{n}y_{k_n} \nonumber \\
= & \ b_1u_{k-1} + \ldots b_m u_{k-m} + w_k + c_1 w_{k-1} +\ldots c_p w_{k-p}, \label{eq:1}
\end{align}
where $u_k$ is the exogenous input, $y_k$ is the measured output, $w_k$ is the process noise, $n, m, p$ are the parameter dimensions (orders) which are assumed to be known, $a_i, b_i, c_i$ are system parameters which are constant but unknown. The process noise is assumed to be Gaussian white noise with zero mean and variance $\sigma^2$ which is also unknown. The exogenous input is known and assumed to be stationary and independent of the process noise. The system parameter vector will be denoted by $\theta^{\star} = [a_1 \ldots a_n\  b_1  \ldots b_m\ c_1 \ldots  c_p]^T$. The time index $k$ is allowed to range from $-\infty$ to $+\infty$. 

Denoting the delay operator by $z^{-1}$, the system model (\ref{eq:1}) can be rewritten as 
\begin{align}
a(z)y_k &=b(z)u_k + c(z) w_k, \label{eq:2}
\end{align}
where $a(z) = 1+a_1z^{-1}+\ldots + a_nz^{-n}$, $b(z)=b_1z^{-1}+\ldots b_mz^{-m}$ and $c(z)=1+c_1z^{-1}+\ldots c_pz^{-p}$. It is further assumed that $c(z)$ is stable (i.e., with all their zeros strictly inside the unit circle) and that $a(z), b(z)$ and $c(z)$ do not have a common factor. 

\begin{lem}\label{lem:0}
Under the assumption that $n\ge m$ and $n\ge p$, the observable-canonical state-space realisation of (\ref{eq:2}) is given by 
\begin{align}
x_{k+1} &=Ax_k+ B_1u_k+B_2w_k \nonumber \\
&=\hspace{-1mm} \left [ \begin{array}{cccc}
0 & \ldots & 0 & -a_n \\ 1 & \ddots & \vdots & -a_{n-1}\\
\ & \ddots & 0 & \vdots \\ 0 & \ldots & 1 & -a_1\end{array}\right ]\hspace{-1mm} x_k + \hspace{-1mm} \left [\begin{array}{c} 0 \\ b_m \\ \vdots \\ b_1 \end{array}\right ]\hspace{-1mm} u_k + \hspace{-1mm} \left [\begin{array}{c} \tilde{c}_n \\  \tilde{c}_{n-1} \\ \vdots \\ \tilde{c}_1 \end{array}\right ]\hspace{-1mm} w_k\nonumber \\
y_k & = Cx_k+w_k = [0 \ \ldots \ 0 \ 1]x_k + w_k, \label{eq:3}
\end{align}
where $x_k$ is the state of the system, and $\tilde{c}_i = c_i - a_i, i=1, 2, \ldots, n$ with the extended $c_n=\ldots =c_{p+1} =0$. 
\end{lem}

See Appendix A for proof. 

\subsection{Reinforcement Learning}
Reinforcement learning (RL) is an iconic tool in machine learning with huge success in applications \cite{Silver} and  has deep connections with the control theory \cite{Bertsekas}.  Consider a system
\begin{align}
x_{k+1} &= f(x_k,u_k,w_k) \nonumber \\
z_k &= g(x_k,u_k,w_k), \label{eq:4}
\end{align}
where $x_k$ is the state, $u_k$ is the control input, $w_k$ is the process noise, $z_k$ is the output known as the cost, $f(\cdot)$ and $g(\cdot)$ are unknown functions. Under the assumption that both the state and output are measurable, the aim of RL is to design an optimal control law $u_k = \pi(x_k)$ such that the following total cost  $J_k$ is minimised:
\begin{align}
J_k &=\mathbb{E}[ \sum_{t=0}^{\infty} \gamma^t z_{k+t}], \label{eq:5}
\end{align} 
where $0<\gamma<1$ is a forgetting factor. In the standard RL terminology, control is called {\em action}, control law is called {\em policy}, forgetting factor is called the {\em discount factor}, $-z_k$ is called the {\em reward}, $-J_k$ is called the {\em value function}, and (\ref{eq:5}) is equivalent to {\em maximise} the value function. 

To get around of the difficulty with unknown $f(\cdot)$ and $g(\cdot)$ and unknown structure of feasible policy $\pi(\cdot)$, total cost and policy are parameterised as $J_k^{\theta}$ and $\pi^{\theta}(\cdot)$ with some (high-dimensional) parameter vector $\theta$. These functions are then approximated using neural networks and an iterative algorithm is applied to tune $\theta$, based on the available $x_k$ and $z_k$ sequences (from simulations and/or experiments) such that $J_k^{\theta}$ is minimised. It is worth noting that apart from some simple cases, RL represents a learning paradigm rather than a guarantee for optimal policies. Important cases where the optimal policy is guaranteed include 1) Markov decision process (MDP) with finite numbers of states and actions \cite{Silver}; 2) linear quadratic regulation (LQR) for state feedback control of linear systems \cite{Lewis}.   

Two types of iterative algorithms are most commonly used in RL: {\em policy iteration} (PI) and {\em value iteration} (VI).  PI aims to improve the policy after each iteration whereas VI focuses on improving the value function. The {\bf key difference} between PI and VI is the following: In PI, a single policy (known as {\em on-policy}) is used in every time step, whereas in VI, different policies (known as {\em off-policy}) can be used in different time steps. This is illustrated in Fig.~\ref{fig:1} below. This seemingly subtle difference has a profound influence on the efficiency and effectiveness of the algorithm. Namely, in PI, a complete evaluation of a new policy needs to be performed before the next iteration, whereas in VI, past value functions evaluated based on old policies can be used in evaluating the new policy, making VI a much more popular choice in RL.  

\begin{figure}[ht]
\begin{picture}(253,100)
\put(0,1){\line(1,0){253}}
\put(0,100){\line(1,0){253}}
\put(0,1){\line(0,1){100}}
\put(253,1){\line(0,1){100}}
\put(5,85){Policy Iteration: Same policy $\pi^{(i)}$ is used throughout}
\put(5,40){Value Iteration: Current and past policies are mixed}
\put(3,65){$\ldots$}
\put(26,65){\vector(1,0){28}} 
\put(16,62){$k$}
\put(34,70){$\pi^{(i)}$}
\put(87,65){\vector(1,0){28}} 
\put(60,62){$k+1$}
\put(93,70){$\pi^{(i)}$}
\put(145,65){\vector(1,0){28}} 
\put(119,62){$k+2$}
\put(149,70){$\pi^{(i)}$}
\put(205,65){\vector(1,0){28}} 
\put(177,62){$k+3$}
\put(209,70){$\pi^{(i)}$}
\put(236,65){$\ldots$}
\put(3,15){$\ldots$}
\put(26,15){\vector(1,0){28}} 
\put(16,12){$k$}
\put(34,20){$\pi^{(i)}$}
\put(87,15){\vector(1,0){28}} 
\put(60,12){$k+1$}
\put(90,20){$\pi^{(i-1)}$}
\put(145,15){\vector(1,0){28}} 
\put(117,12){$k+2$}
\put(148,20){$\pi^{(i-2)}$}
\put(205,15){\vector(1,0){28}} 
\put(177,12){$k+3$}
\put(206,20){$\pi^{(i-3)}$}
\put(236,15){$\ldots$}
\end{picture}
  \caption{Illustration of Policy Iteration and Value Iteration}\label{fig:1}
\end{figure}
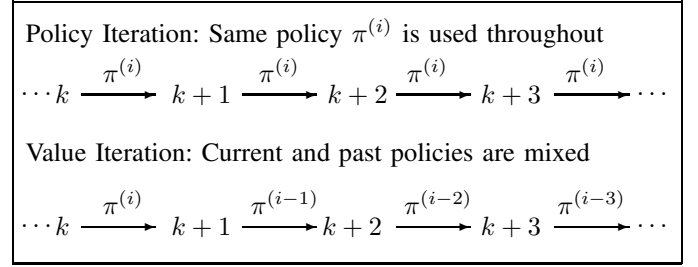 \vspace{-3mm}

 \subsection{RL Formulation of System Identification}

We now formulate the system identification problem as a reinforcement learning problem. Let 
\begin{align} 
\hat{y}_k = \pi(y_{<k}, u_{<k}), \label{eq:6}
\end{align}
be a (one-step-ahead) predictor of $y_k$, where $y_{<k} = [y_{k-1}, y_{k-2}, \ldots ]$ and $u_{<k}$ is similarly defined. The prediction error is given by 
\begin{align}
e_k = y_k - \hat{y}_k. \label{eq:7}
\end{align}
The total cost is defined to be  
\begin{align}
J_k = \mathbb{E} [ \sum_{t=0}^{\infty} \gamma^t e_{k-t}^2 ] \label{eq:8}
\end{align}
for some discount factor $0<\gamma<1$. Notice that this sequence goes {\bf backwards} in time, and that the initial state is not present because the sequence of $y_k$ starts from $k=-\infty$. The RL problem is to find the optimal policy (i.e., predictor) $\pi$ such that $J_k$ is minimised. We will show later that this formulation coincides with the classical system identification problem. 

\subsection{RL Formulation of Optimal State Estimation}

State estimation without a system model has a unique difficulty due to infinite choices of state coordinates. Therefore, the state estimation problem not only needs to provide an optimal state estimate, but also to specify the system structure. Mathematically, we need to determine the following model:
\begin{align}
\pi: \ \ \ \hat{x}_{k+1} &= \hat{f}(\hat{x}_k, y_k, u_k)\nonumber \\
\hat{y}_k& = \hat{g}(\hat{x}_k) \label{eq:9}
\end{align}
where $\hat{x}_k$ represents the estimated state, $\hat{f}(\cdot)$ and $\hat{g}$ are the unknown functions (i.e., structure and parameters). Collectively,  $\hat{f}(\cdot)$ and $\hat{g}(\cdot)$ constitute the policy to be optimised.  A ``simple" choice for the estimated state is $\hat{x}_k=\mathrm{col}[y_{<k}, u_{<k}]$, but this is not desirable because its dimension is infinite. It is natural that we want $\hat{x}_k$ to have a fixed finite dimension.

The RL formulation for model-free state estimation is to find the optimal $\pi$ in (\ref{eq:9}) such that the total cost $J_k$ in (\ref{eq:8}) is minimised. Again, we will show later that this formulation is consistent with the classical Kalman filtering problem. 

\section{System Identification}

This section solves the RL problem for system identification. 

We first make a simple observation that the discount factor does not play any role and that the problem formulation (\ref{eq:8}) can be simplified. 

\begin{lem}\label{lem:1}
For any given (stationary) policy $\pi$ in (\ref{eq:6}), the total cost $J_k$ in (\ref{eq:8}) can be simplified to 
\begin{align} 
J_k &= \frac{1}{1-\gamma} \mathbb{E}[e_k^2]. \label{eq:10}
\end{align}
\end{lem}

\begin{IEEEproof}
The result follows from the stationarity of the system model (\ref{eq:1}) and that of the policy. That is, 
$\mathbb{E}[e_k^2]$ is independent of $k$.  Hence, $J_k = \mathbb{E}[e_k^2]\sum_{t=0}^{\infty} \gamma^t$, giving (\ref{eq:10}). 
\end{IEEEproof}

The result above indicates that we effectively minimise the squared prediction error, for which  the following holds.
\begin{lem}\label{lem:2}
Suppose, for any $k$, $w_k$ is independent of $u_{k-i}$ for any $i=1, \ldots, m$. Then, the optimal policy $\pi^{\star}$ of (\ref{eq:6}) that minimises $\mathbb{E}[J_k]$ is given by 
\begin{align}
\hat{y}_k =& -a_1y_{k-1} - \ldots -a_ny_{k-n} + b_1u_{k-1} + \ldots + b_mu_{k-m}\nonumber \\
&  +c_1e_{k-1} +\ldots + c_pe_{k-p} \label{eq:11}
\end{align}
with $e_{k-i}=y_{k-i}-\hat{y}_{k-i}$ defined recursively. The corresponding minimum is given by
\begin{align}
\min_{\pi} \mathbb{E}[J_k] &= \frac{1}{1-\gamma}\sigma^2. \label{eq:12}
\end{align}
\end{lem}

\begin{IEEEproof}
Firstly, it is obvious from (\ref{eq:1}) and (\ref{eq:6})-(\ref{eq:7}) that $e_k$ can be rewritten as 
\begin{align*}
e_k = w_k + \tilde{e}_k
\end{align*}
where $\tilde{e}_k$ a function of $u_{<k}, y_{<k}$ and $w_{<k}$, hence independent of $w_k$. It is clear that $\mathbb{E}[e_k^2]\ge \mathbb{E}[w_k^2] = \sigma^2$. By taking $\hat{y}_k$ as in (\ref{eq:11}), we get 
\begin{align*}
e_k + c_1 e_{k-1} + \ldots c_p e_{k-p} = w_k + c_1 w_{k-1} + \ldots c_p w_{k-p},
\end{align*} 
i.e., $e_k$ and $w_k$ have the same power spectrum, hence $\mathbb{E}[e_k^2]= \mathbb{E}[w_k^2] = \sigma^2$, confirming the optimality of (\ref{eq:11}). Finally, (\ref{eq:12}) is obtained by using Lemma~\ref{lem:1}. (We note that the result for minimum $\mathbb{E}[e_k^2]$ is consistent with \cite{Ljung}.)
\end{IEEEproof}

With Lemma~\ref{lem:2}, we can take the policy structure to be 
\begin{align}
\hat{y}_k(\theta) =& -\hat{a}_1y_{k-1} - \ldots -\hat{a}_ny_{k-n}\hspace{-1mm}  + \hat{b}_1u_{k-1}\hspace{-1mm}  + \ldots + \hat{b}_mu_{k-m}\nonumber \\
&  +\hat{c}_1e_{k-1}(\theta) +\ldots + \hat{c}_pe_{k-p}(\theta) \label{eq:13}\\
e_k(\theta) &= y_k -\hat{y}_k(\theta), \label{eq:14}
\end{align}
with $\theta= [\hat{a}_1\ \ldots \ \hat{a}_n\  \hat{b}_1\  \ldots\ \hat{b}_m\ \hat{c}_1\ \ldots\  \hat{c}_p]^T$. 

The most popular method for ARMAX estimation is the so-called {\em pseudo-linear regression} (PLR) method~\cite{Ljung,Soderstrom}. Defining the pseudo-linear regressor as
\begin{align}
\varphi_k(\theta) =& [-y_{k-1}\ \ldots\ -y_{k-n}\ u_{k-1}\ \ldots\ u_{k-m}\nonumber \\
& \ e_{k-1}(\theta)\ \ldots \ e_{k-p}(\theta)]^T, \label{eq:15}
\end{align}
then
\begin{align}
\hat{y}_k(\theta) &= \varphi_k^T(\theta)\theta; \ \ e_k(\theta)=y_k-\varphi_k^T(\theta)\theta. \label{eq:16}
\end{align}  

The PRL estimate of $\theta$ is computed by solving 
\begin{align}
\mathbb{E}[\varphi_k(\theta)(y_k-\varphi_k^T(\theta)\theta)] = 0. \label{eq:17}
\end{align}
This is typically done recursively (known as {\em bootstrapping} method in the system identification literature): Starting from some initial estimate $\theta^{(0)}$, then for each $i=1,2,\ldots$, solve $\theta^{(i)}$ using
\begin{align} 
\mathbb{E}[\varphi_k(\theta^{(i-1)})(y_k-\varphi_k^T(\theta^{(i-1)})\theta^{(i)})] = 0, \label{eq:18}
\end{align}
which is a {\em repeated least-squares} problem. 

The PRL method is also often combined with the {\em instrumental variable} method, where the first term $\varphi_k(\theta)$ in (\ref{eq:17}) is replaced with an instrumental variable (vector) $\zeta_k(\theta)$ which is designed to be uncorrelated with $e_k(\theta)$. 

The convergence properties of the bootstrapping and the instrumental variable method depend on many factors; see \cite{Soderstrom,Ljung} for detailed analysis.  We emphasise two key observations: 
\begin{enumerate}
\item Global convergence to the optimal solution is not always guaranteed; 
\item  The bootstrapping method above is a form of policy iteration in the viewpoint of RL. 
\end{enumerate}
We will see below that by using a value iteration method in combination with the instrumental variable method, a globally convergent algorithm can be derived for ARMAX estimation. We will first study off-line identification before giving an on-line algorithm. 

\subsection{Off-line Identification of MA Models}
We first consider the case of MA models, as this is the stumbling block in system identification, causing the regressor (\ref{eq:15}) to depend on $\theta$. The MA model is given by 
\begin{align}
y_k &= w_k + c_1w_{k-1}+\ldots c_p w_{k-p} \label{eq:19}
\end{align} 
with the assumption that $c(z)=1+c_1z^{-1} +\ldots + c_pz^{-p}$ is strictly stable. Also, $\theta=[\hat{c}_1 \ \ldots \ \hat{c}_p]^T$ in this case.

Identification of MA models can be traced back at least to \cite{Durbin,Walker}. But earlier methods all require solving difficult nonlinear equations. In \cite{Ljung} (p.~337), a two-step, non-iterative method is provided: Step 1 estimates a high-order AR model to approximate the MA model; Step 2 uses the prediction errors (known as innovations) from the AR model as an estimate of the past process noise and estimate the MA parameters using the least-squares method. This method requires heavy computation for the first step due to the use of a high-order AR model and gives only an approximate solution.  Alternatively, the PRL method can be used to reduce complexity, but there is no theoretical guarantee for an optimal solution. 

Here we introduce a new algorithm based on an VI method in RL. That is, {\bf we generalise the bootstrapping method (\ref{eq:18}) by allowing the PLR to depend on multiple past estimates of $\theta$}, as illustrated in Fig.~\ref{fig:1}. We first consider an {\em off-line} iterative algorithm before extending it to on-line learning. 

We start estimating $\theta$ from $k=0$. Denote by $\theta^{(k)}, k\ge0$ the $k$-th estimate. We revise (\ref{eq:13})-(\ref{eq:14}) to the following:
\begin{align}
\hat{y}_k(\theta) =&   \hat{c}_1e_{k-1}(\theta^{(k-1)}) +\ldots + \hat{c}_pe_{k-p}(\theta^{(k-p)}) \label{eq:13-1}\\
e_k(\theta) &= y_k -\hat{y}_k(\theta). \label{eq:14-1}
\end{align}
That is, $e_{k-i}(\theta)$ is replaced with $e_{k-i}(\theta^{(k-i)})$. With some abuse of notation, the latter will denoted by $e_k$ if not confusing.

Suppose $y_k$ is measured for all $k<0$. Due to the stationarity of (\ref{eq:19}), we can compute all the autocorrelations $r_y(i)=\mathbb{E}[y_ky_{k-i}]$ using the available measurements prior to $k=0$. Our off-line  iterative algorithm assumes that $r_y(i), i=0,1,\ldots, p$, are available and produces a sequence of $\theta^{(k)}, k\ge0,$ such that $\theta^{(k)}\rightarrow \theta^{\star}$ as $k\rightarrow \infty$.  

Initialise $e_{-1} = \ldots = e_{-p} = 0$ and $\theta^{(-1)}=\ldots = \theta^{(-p)}=0$. For $k=0, 1, \ldots$, solve $\theta^{(k)}$ from
\begin{align}
\min_{\theta} \mathbb{E}[e_k^2] = \mathbb{E}[(y_k - \hat{c}_1 e_{k-1} - \ldots - \hat{c}_p e_{k-p})^2]\label{eq:20}
\end{align}
and construct the resulting $e_k$ using $\theta^{(k)}$.  Differentiating the above results in the orthogonality condition: 
\begin{align}
\mathbb{E}[e_ke_{k-i}] = 0, i =1, \ldots, p.  \label{eq:21}
\end{align}
As we will show later that the orthogonality condition holds recursively, i.e., $\mathbb{E}[e_{k-j}e_{k-j-i}]=0$ for all $j>0$ and $i>0$ as well. This implies that (\ref{eq:21}) can be simplified to 
\begin{align*}
\mathbb{E}[(y_k-\hat{c}_i e_{k-i})e_{k-i}] =0,
\end{align*}
giving the simple solution for $\hat{c}_i$ as 
\begin{align}
c_i^{(k)} = \left \{ \begin{array}{ll}\rho_k(i)/\mathbb{E}[e_{k-i}^2]  & \mathrm{if\ } \mathbb{E}[e_{k-i}^2]>0\\0 & \mathrm{otherwise}
\end{array}\right., \label{eq:22}
\end{align}
where $\rho_k(i) = \mathbb{E}[y_ke_{k-i}]$.

The resulting $\mathbb{E}[e_k^2]$ is given by 
\begin{align}
\mathbb{E}[e_k^2] = r_y(0) - (c_1^{(k)})^2 \mathbb{E}[e_{k-1}^2] - \ldots (c_p^{(k)})^2 \mathbb{E}[e_{k-p}^2]. \label{eq:23}
\end{align}
Note that, by construction, $e_k$ explicitly depends on $\theta^{(k)}$ but implicitly depends on $\theta^{(k-1)}, \theta^{(k-2)}, \ldots$ because of $e_{k-1}, \ldots$, hence the method above is an VI method in the RL framework. 

The computation of (\ref{eq:22}) involves $\rho_k(i)$, which can also be easily updated.  Indeed, for $i=1,\ldots, p$,
\begin{align*}
\rho_k(i)=& \mathbb{E}[y_ke_{k-i}] \\
=& \mathbb{E}[y_k(y_{k-i}-c_1^{(k-i)} e_{k-i-1} - \ldots -c_p^{(k-i)} e_{k-i-p})] \\
=& r_y(i) - c_1^{(k-i)}\rho_k(i+1) - \ldots -c_p^{(k-i)}\rho_k(i+p).
\end{align*}
From (\ref{eq:19}), $\rho_k(i)=\mathbb{E}[y_ke_{k-j}]=0$ for $j>p$. Therefore, 
\begin{align}
\hspace{-2mm}\left [\hspace{-1mm} \begin{array}{cccc} 1 & c_1^{(k-1)} & \ldots &  c_{p-1}^{(k-1)} \hspace{-1mm}\\
0 & 1 & c_1^{(k-2)} & \vdots  \\
\vdots & \ddots & 1& c_1^{(k-p+1)}\hspace{-2mm} \\
0 & \ldots &  0 & 1\end{array}\right ]\hspace{-1mm}\left [ 
\begin{array}{c} \rho_k(1) \\ \rho_k(2)\\ \vdots \\ \rho_k(p)\end{array}\hspace{-1mm}\right ]\hspace{-1mm} =\hspace{-1mm} \left [\hspace{-1mm} \begin{array}{c} r_y(1) \\ r_y(2)\\ \vdots \\ r_y(p)\end{array}\hspace{-1mm}\right ] \label{eq:24}
\end{align}
which can be easily computed due to the triangular structure. 

We have the following result for convergence. 
\begin{thm}\label{thm:1}
The VI method above has two properties:
\begin{itemize}
\item (Orthogonality:) $\mathbb{E}[e_ke_{k-i}] =0$ for all $k\ge i$ and $i>0$; 
\item (Convergence and Consistency:) $\theta^{(k)}\rightarrow \theta^{\star}$ as $k\rightarrow \infty$. 
\end{itemize}
\end{thm}

\begin{IEEEproof}
Take any $k\ge0$ and $i>0$. The orthogonality condition for $i\le p$ was given in (\ref{eq:21}). Now consider $i=p+1$,  
\begin{align*}
\mathbb{E}[e_ke_{k-i}] &= \mathbb{E}[(y_k-c_1^{(k)}e_{k-1}- \ldots -c_p^{(k)} e_{k-p} )e_{k-i}]
\end{align*}
The first term $\mathbb{E}[y_ke_{k-i}]=0$ due to $i>p$. Thus, $\mathbb{E}[e_ke_{k-i}]=0$ because $\mathbb{E}[e_{k-j}e_{k-i}]=0$ for all $j=1,\ldots, p$ due to $i=p+1$. This process can be repeated for $i=p+2, p+3, \ldots$. Hence, the orthogonality condition holds for all $i>0$. 

To show convergence and consistency, we note that the sequences $\{e_0, e_1, \ldots e_{k-1}\}$ and $\{y_0, y_1, \ldots, y_{k-1}\}$ form a linear invertible mapping. Let $\pi_k^{\star}$ be the optimal function in (\ref{eq:6}), linear or nonlinear, such that $\mathbb{E}[(y_k-\hat{y}_k)^2]$ is minimised. Due to the assumption that $w_k$ is a Gaussian white noise, it is well-known \cite{Anderson} that the optimal $\pi_k^{\star}$ is a linear mapping. Due to the invertibility above, the optimal $\hat{y}_k$ can be represented as the following linear mapping:
\begin{align*}
\hat{y}_k &= \delta_1 e_{k-1} + \ldots \delta_pe_{k-p} +\delta_{p+1}e_{k-p-1}+ \ldots+\delta_k e_0 
\end{align*}
and the optimal $\delta_i$ can be solved by minimising $\mathbb{E}[(y_k-\hat{y}_k)^2]$. 
Due to the orthogonality peroperty of $e_k$ and the fact that $y_k$ is orthogonal to $e_{k-i}$ for $i>p$, it is easy to see that $\delta_i=0$ for any $i>p$ and, for any $i=1, 2, \ldots, p$, $\delta_i$ are the same as $\hat{c}_i$ in (\ref{eq:22}). That is, the optimal $\hat{y}_k$ is given by
\begin{align}
\hat{y}_k &= \hat{c}_1 e_{k-1} + \ldots + \hat{c}_pe_{k-p}  \label{eq:25}
\end{align}
with $[\hat{c}_1 \ \ldots \hat{c}_p]=\theta^{(k)}$. 

On the other hand, it is also well known \cite{Anderson} that, as $k\rightarrow \infty$, the stability of $c(z)$ and stationarity of (\ref{eq:20}) implies that the optimal $\hat{y}_k$ is such that 
\begin{align*}
e_k = y_k - \hat{y}_k \rightarrow c^{-1}(z)y_k = w_k
\end{align*}
That is,  $c(z)e_k \rightarrow y_k$, i.e.,  
\begin{align*}
e_k &\rightarrow  -c_1 e_{k-1} - \ldots - c_p e_{k-p} + y_k 
\end{align*}
\begin{align*}
\hat{y}_k &\rightarrow  c_1 e_{k-1} + \ldots c_p e_{k-p} 
\end{align*}
Comparing this to (\ref{eq:25}), we see that $\theta^{(k)}\rightarrow \theta^{\star}$ as $k\rightarrow \infty$. 
\end{IEEEproof}

\subsection{Off-line Identification of ARMAX Model}

Now let us return to the ARMAX model (\ref{eq:1}). Using the policy structure (\ref{eq:13})-(\ref{eq:17}), the task is to solve $\theta$ to minimise
\begin{align}
\mathbb{E}[e_k^2(\theta)] = \mathbb{E}[(y_k -\varphi_k(\theta)^T\theta)^2]. \label{eq:26}
\end{align}

However, the coupling between the ARX part and MA part of the model makes it difficult to minimise (\ref{eq:26}) directly. To get around this difficulty, we can first use a classical instrumental variable method in system identification to estimate the ARX part and then the proposed VI method to estimate the MA model \cite{Ljung,Soderstrom}.

Define the instrumental variable (vector) as
\begin{align}
\zeta_k &= F(z)[-y_{k-p-1}\ \ldots -y_{k-p-n} \ u_{k-1} \ \ldots \ u_{k-m}]^T, \label{eq:27}
\end{align}
where $F(z)$ is a causal linear filter with both $F(z)$ and $F^{-1}(z)$ being stable. In particular, we can take $F(z)=1$. 
Using (\ref{eq:1}) and properties of $u_k$ and $w_k$, we get 
\begin{align}
\mathbb{E}[\zeta_k (&y_k+a_1y_{k-1}+\ldots +a_ny_{k-n} \nonumber \\
&- b_1u_{k-1} - \ldots -b_mu_{k-m})]=0. \label{eq:28}
\end{align}
This gives $(n+m)$ linear equations: 
\begin{align}
R\tilde{\theta} = \mathbf{r} \label{eq:29}
\end{align}
with $\tilde{\theta}=\mathrm{col}\{a, b\}, R=\mathbb{E}[\zeta_k\tilde{\varphi}_k^T],  \mathbf{r}=\mathbb{E}[\zeta_ky_k]$ and  $\tilde{\varphi}_k = [-y_{k-1} \ \ldots \ -y_{k-n} \ u_{k-1} \ \ldots \ u_{k-m}]^T$. This allows us to solve $a$ and $b$ under the mild persistent excitation condition of nonsingular $R$ \cite{Ljung,Soderstrom}. 

For the case of ARMA models (with $b=0$), if $F(z)=I$, then $\zeta_k =[-y_{k-p-1} \ \ldots \ -y_{k-p-n}]^T$ and (\ref{eq:29}) reduces to 
\begin{align}
R_y a &= \mathbf{r}_y \label{eq:30}
\end{align}
with $R_y = \mathbb{E}[\zeta_k\zeta_{k+p}^T]$ and $\mathbf{r}_y = \mathbb{E}[\zeta_ky_k]$,
and we have the following result. 
\begin{prop}\label{prop:1}
For the case of an ARMA model, $R_y$ is nonsingular if $a_n\ne0$ and $c(z)/a(z)$ is a minimal realisation (i.e., $a(z)$ is not degenerate in its order and there is no zero-pole cancellation between $c(z)$ and $a(z)$).  
\end{prop}
\begin{IEEEproof}
See Appendix~B.
\end{IEEEproof}

After the ARX part of the model is identified, we define
\begin{align}
\tilde{y}_k &=y_k-\tilde{\varphi}_k^T\tilde{\theta}. \label{eq:31}
\end{align}
Then the new MA model 
\begin{align}
\tilde{y}_k &= w_k + c_1w_{k-1} + \ldots c_p w_{k-p} \label{eq:32}
\end{align}
can be identified by the value iteration method for MA models. 
 
\subsection{On-line Identification of ARMAX Models}

In order to obtain an on-line identification method for ARMAX models, we need to convert the instrumental variable method for the ARX part into a recursive algorithm and combine it with a recursive algorithm of the value iteration method for the MA model. 

We do the conversion for the ARX part first. At each time instant $k=1,2,\ldots$, we replace (\ref{eq:29}) with 
\begin{align}
R^{(k)}\tilde{\theta}^{(k)} & =\mathbf{r}^{(k)}, \label{eq:33}
\end{align}
by approximating expectations with empirical averages, i.e., 
\begin{align}
R^{(k)}&= \frac{1}{k+1} \sum_{t=0}^k \zeta_t\tilde{\varphi}_t^T= \frac{k}{k+1} R^{(k-1)} + 
\frac{1}{k+1}\zeta_k\tilde{\varphi}_k^T, \label{eq:34} \\
\mathbf{r}^{(k)}&= \frac{1}{k+1} \sum_{t=0}^k \zeta_ty_t = \frac{k}{k+1} \mathbf{r}^{(k-1)} + 
\frac{1}{k+1}\zeta_ky_k.  \label{eq:35}
\end{align}
Denoting $P^{(k)} = (R^{(k)})^{-1}$, it is standard \cite{Ljung} to obtain the recursive solution to (\ref{eq:33}) as below.
\begin{prop}\label{prop:2}
The solution to (\ref{eq:33}) has the following recursion for $k\ge1$:
\begin{align} 
\tilde{\theta}^{(k)}& = \tilde{\theta}^{(k-1)} + \frac{1}{k}P^{(k-1)}\zeta_k\gamma_k^{-1}(y_k-\tilde{\varphi}_k^T\tilde{\theta}^{(k-1)}),  \label{eq:36}\\
P^{(k)}&=\left [I - \frac{1}{k}P^{(k-1)}\zeta_k\gamma_k^{-1}\tilde{\varphi}_k^T\right ]\frac{k+1}{k}P^{(k-1)} \label{eq:37}
\end{align}
where
\begin{align}
\gamma_k &= 1+\frac{1}{k}\tilde{\varphi}_k^TP^{(k-1)}\zeta_k \label{eq:38}
\end{align}
Moreover, $\tilde{\theta}^{(k)}\rightarrow \tilde{\theta}^{\star}$ (the true value of col$\{a,b\}$) and $R^{(k)}\rightarrow R$ as $k\rightarrow \infty$ with probability 1, provided that $R$ is nonsingular.  
\end{prop}
\begin{IEEEproof}
The recursion (\ref{eq:37}) is obtained by applying the well-known matrix inversion lemma to (\ref{eq:34}). Then, (\ref{eq:36}) is obtained by applying (\ref{eq:35}) and (\ref{eq:37}) to solving (\ref{eq:33}). Also, $R^{(k)}\rightarrow  R$ and $\mathbf{r}^{(k)}\rightarrow \mathbf{r}$ (with probability 1) owing to the stationarity of the system, and $\tilde{\theta}^{(k)}\rightarrow \tilde{\theta}^{\star}$ (with probability 1) because $R$ is nonsingular. 
\end{IEEEproof}

Next, we convert the MA part. First, we revise $\tilde{y}_k$ to 
\begin{align}
\tilde{y}_k &=y_k-\tilde{\varphi}_k^T \tilde{\theta}^{(k)}, \label{eq:39}
\end{align}
which converges to (\ref{eq:31}) as $k\rightarrow \infty$.  

Secondly, we replace $r_y(i), i=0,1,\ldots, p,$ with empirical averages, i.e., 
\begin{align}
r_y^{(k)}(i) &= \frac{1}{k+1}\sum_{t=0}^k \tilde{y}_t\tilde{y}_{t-i} \nonumber \\
 &= \frac{k}{k+1}r_y^{(k-1)}(i) + \frac{1}{k+1}\tilde{y}_k\tilde{y}_{k-i}. \label{eq:40}
\end{align}
Again, $r_y^{(k)}(i)\rightarrow r_y(i)$ as $k\rightarrow \infty$. 

Thirdly, using $r_y^{(k)}(i)$ above, we modify (\ref{eq:24}) to
\begin{align}
\left [\hspace{-1mm} \begin{array}{cccc} 1 & c_1^{(k-1)} & \ldots &  c_{p-1}^{(k-1)} \\
0 & 1 & c_1^{(k-2)} & \vdots  \\
\vdots & \ddots & 1& c_1^{(k-p+1)} \\
0 & \ldots &  0 & 1\end{array}\right ]\hspace{-1mm}\left [ 
\begin{array}{c} \rho_k(1) \\ \rho_k(2)\\ \vdots \\ \rho_k(p)]\end{array}\hspace{-1mm}\right ]\hspace{-1mm} =\hspace{-1mm} \left [\hspace{-1mm} \begin{array}{c} r_y^{(k)}(1) \\ r_y^{(k)}(2)\\ \vdots \\ r_y^{(k)}(p)\end{array}\hspace{-1mm}\right ] \label{eq:43}
\end{align}

Finally,  we replace $\mathbb{E}[e_k^2]$ with $\epsilon_k^2$ and modify (\ref{eq:22})-(\ref{eq:23}) as  
\begin{align}
c_i^{(k)} &= \left \{ \begin{array}{ll}\rho_k(i)/\epsilon_{k-i}^2  & \mathrm{if\ } \epsilon_{k-i}^2>0\\0 & \mathrm{otherwise}
\end{array}\right., i=1,2,\ldots, p,\label{eq:41}\\
\epsilon_k^2 &= r_y^{(k)}(0) - (c_1^{(k)})^2 \epsilon_{k-1}^2 - \ldots (c_p^{(k)})^2 \epsilon_{k-p}^2; \label{eq:42}
\end{align}

The resulting on-line  algorithm is summarised below. 
\begin{algorithm}[ht] 
\protect\protect\protect\protect\protect\protect\protect\caption{(On-line Identification for ARMAX Models)}
\label{alg:1} \begin{itemize}
\item \textbf{Initialisation:} 
\begin{itemize}
\item Set $\epsilon_{-i}^2=0, i=1,2,\ldots p-1$ and $\epsilon_0^2=y_0^2$;
\item Set $r_y^{(0)}(i) = 0, i=1,2,\ldots,p$ and $r_y^{(0)}(0)=y_0^2$; 
\item Set $\mathrm{col}\{\tilde{\theta}^{(0)}, c_1^{(0)},\ldots, c_p^{(0)}\}=0$;
\item Set $P^{(0)}=p_0 I$ for any (large) $p_0>0$.
\end{itemize}
\item \textbf{Main loop:} At iteration $k=1,2,\cdots$, 
\begin{itemize}
\item Compute $\tilde{\theta}^{(k)}$ and $P^{(k)}$ using (\ref{eq:36})-(\ref{eq:37});
\item Compute $r_y^{(k)}(i), i=0,1,\ldots,p,$ using (\ref{eq:40});
\item Compute $\rho_k(i), i=1,2,\ldots, p,$ using (\ref{eq:43});
\item Compute $c_i^{(k)}, i=1,2,\ldots,p,$ using (\ref{eq:41});
\item Compute $\epsilon_k^2$ using (\ref{eq:42});
\end{itemize}
\end{itemize}
\end{algorithm}

We have the following result.
\begin{thm}\label{thm:2}
Under the persistent excitation condition $R>0$, Algorithm~\ref{alg:1} has the following properties as $k\rightarrow\infty$:
\begin{itemize}
\item $\epsilon_k^2\rightarrow \mathbb{E}[e_k^2]$ with probability 1;
\item $\theta^{(k)}\rightarrow \theta^{\star}$ with probability 1.
\end{itemize}
\end{thm}

\begin{IEEEproof}
The proof follows directly from Theorem~\ref{thm:1}, Proposition~\ref{prop:2}, and  $r_y^{(k)}(i)\rightarrow r_y(i)$ with probability 1 for all $i$.
\end{IEEEproof}

 \section{Model-based State Estimation}

\subsection{Optimal State Estimation for a Known Model}
Consider the following state-space model:
\begin{align}
x_{k+1}&=Ax_k + B_1u_k + B_2w_k \nonumber \\
y_k &=Cx_k +v_k, \label{eq:201}
\end{align}
where $x_k$ is the state, $u_k$ is the known input, $w_k$ is the process noise, $v_k$ is the measurement noise, $\{(w_k,\ v_k)\}$ is zero-mean Gaussian noise with 
\begin{align}
\mathbb{E}\left \{ \left [ \begin{array}{c} w_k \\ v_k \end{array}\right ] [w_l^T \ v_l^T]\right \} &= \left [ \begin{array}{cc} Q & S \\ S^T & R\end{array}\right ] \delta_{kl}, \ \ k,l\in \mathbb{R}. \label{eq:202}
\end{align}
When the system model is known, the steady-state Kalman filter of (\ref{eq:201}) is given by \cite{Anderson}
\begin{align}
\hat{x}_{k+1} & = A\hat{x}_k + B_1u_k+ L(y_k-C\hat{x}_k)  \label{eq:203}
\end{align} 
with the optimal observer gain $L$ given by \cite{Anderson} (Section 5.4)
\begin{align}
L&=(A\Sigma C^T+B_2S)(C\Sigma C^T+R)^{-1} \label{eq:204} \\
\Sigma&= A\Sigma A^T - (A\Sigma C^T+B_2S)(C\Sigma C^T+R)^{-1}\nonumber \\
&\hspace{18mm} \cdot(A\Sigma C^T+B_2S)^T+B_2QB_2^T. \label{eq:205}
\end{align}
In the above, $\Sigma=\mathbb{E}[(x_k-\hat{x}_k)(x_k-\hat{x}_k)^T]$ is the steady-state state estimation error covariance, and (\ref{eq:204}) is an algebraic Riccati equation (ARE). 

\subsection{Pitfall for Model-Free State Estimation}
\begin{ex}\label{ex:1}
Consider the scalar sequence $\{y_k\}$:
\begin{align} 
y_k &= w_{k-1}+w_k + \mu_k = (1+z^{-1})w_k+\mu_k \label{eq:210-1}
\end{align}
which has the following state-space realisation:
\begin{align}
x_{k+1} & = w_k \nonumber  \\
y_k &= x_k + w_k + \mu_k \label{eq:210}
\end{align} 
where $w_k$ and $\mu_k$ are independent zero-mean Gaussian white noises with variance equal to 1. Comparing with (\ref{eq:201})-(\ref{eq:202}), we verify that $v_k = w_k+\mu_k, A=1, B_1=0, B_2=1, C=1,  Q=1, R=2, S=1$. The state estimator (\ref{eq:203}) becomes 
\begin{align}
\hat{x}_{k+1}&= L(y_k-\hat{x}_k). \label{eq:211}
\end{align}
Solving (\ref{eq:205}) gives 
\begin{align*}
\Sigma&= \Sigma - (\Sigma+1)^2(\Sigma+2)^{-1}+1
\end{align*}
resulting in $\Sigma= (\sqrt{5}-1)/2\approx 0.618$ and $L\approx 0.618$.

On the other hand, the spectrum of $y_k$ in (\ref{eq:210-1}) is 
\begin{align*}
S_y&= (1+z^{-1})(1+z) + 1 = \alpha(1+ \alpha^{-1}z^{-1})(1 + \alpha^{-1}z)
\end{align*}
with $\alpha = 0.5(3+\sqrt{5})$.
Now consider an alternative state-space realisation:
\begin{align}
\mathbf{x}_{k+1} & = \mathbf{w}_k \nonumber \\
y_k &= \alpha^{-1}\mathbf{x}_k + \mathbf{w}_k \label{eq:212}
\end{align}
with $\mathbf{w}_k$ being a zero-mean Gaussian white noise with variance of $\alpha$. For (\ref{eq:212}), the counterpart of $(Q,R,S)$ is given by $\mathbf{Q}=\mathbf{R}=\mathbf{S}=\alpha$. The optimal state estimator is given by 
\begin{align}
\hat{\mathbf{x}}_{k+1} & = \mathbf{L}(y_k-\hat{\mathbf{x}}_k).  \label{eq:213}
\end{align} 
Solving (\ref{eq:205}) for this estimator gives $\mathbf{L} = 1$, and the corresponding steady-state state estimation error covariance $\Sigma = 0$. That is, in steady state, $\mathbf{x}_k$ can be {\em perfectly} predicted by $y_{<k}$!  

Since the noises $w_k$ and $\mu_k$ are not directly measurable, the state-space representation (\ref{eq:201}) is indistinguishable from (\ref{eq:212}). 

We see from this example that different state-space realisations can result in vastly different state estimation results, which is a unique feature for the state estimation problem when the state-space model is not specified!
\end{ex}

\subsection{State-Space Realisation for ARMAX Models}
Motivated by Example~\ref{ex:1}, we see that different state-space realisations may result in vastly different state estimation errors. Here, we present a state-space realisation that has zero optimal state estimation error in steady state. 

Our chosen state-space realisation for (\ref{eq:1}) is (\ref{eq:3}), for which we have the following result. 

\begin{thm}\label{thm:3}
Suppose the system model (\ref{eq:2}) is such that 1) $n\ge m$ and $n\ge p$; and 2) $c(z)$ is stable. Then, the optimal state estimator (\ref{eq:203}) for the state-space realisation (\ref{eq:3}) has the observer gain 
\begin{align}
L&=B_2 = [\tilde{c}_n \ \ldots \tilde{c}_1]^T \label{eq:214}
\end{align}
and its associated state estimation error is zero with probability 1 in steady state, i.e., $\Sigma=0$. 
\end{thm}

\begin{IEEEproof}
Comparing the realisation (\ref{eq:3}) with (\ref{eq:201})-(\ref{eq:202}), it is clear that $Q=S=R = \mathbb{E}[w_kw_k^T] = \sigma^2$ for (\ref{eq:3}). Using the state estimator in (\ref{eq:203}) and defining the estimation error $\varepsilon_k = x_k-\hat{x}_k$ and its covariance $\Sigma_k=\mathbb{E}[\varepsilon_k\varepsilon_k^T]$, we have
\begin{align*}
\varepsilon_{k+1} &= (A-LC) \varepsilon_k + (B_2-L)w_k\\
 \Sigma_{k+1} &= (A-LC)\Sigma_k (A-LC)^T+\sigma^2 (B_2-L)(B_2-L)^T.
 \end{align*}
It is clear that if $L=B_2$ then, 
\begin{align}
\Sigma_{k+1}&=(A-B_2C)\Sigma_k (A-B_2C)^T.\label{eq:215}
\end{align}
From Lemma~\ref{lem:0}, we have
\begin{align*}
A-B_2C &=
\hspace{-1mm} \left [ \begin{array}{cccc}
0 & \ldots & 0 & -c_n\\ 1 & \ddots & \vdots & -c_{n-1}\\
\ & \ddots & 0 & \vdots \\ 0 & \ldots & 1 & -c_1\end{array}\right ]
\end{align*}
It is easy to verify that 
\begin{align}
\mathrm{det}(I-(A-B_2C)z^{-1}) = c(z), \label{eq:216}
\end{align}
which is assumed to be stable. Hence, $A-B_2C$ is stable. It follows from (\ref{eq:215}) that  $\Sigma_{k}\rightarrow 0$ as $k\rightarrow \infty$. That is, the steady-state estimation error covariance $\Sigma=0$, which is obviously optimal.  Hence,  $L=B_2$ is the optimal observer gain. 
\end{IEEEproof}

\subsection{Model-free State Estimation}
The result in Theorem~\ref{thm:3} shows that with an appropriate choice of the state-space realisation, perfect state estimation can be achieved asymptotically. However, this result requires known parameters for the system. We now show how to achieve something similar without a known model.  

Using Algorithm~\ref{alg:1}, we can build the one-step-ahead prediction $\hat{y}_k$ of $y_k$ and the prediction error $e_k$ as 
\begin{align}
\hat{y}_k =& -a_1^{(k)}y_{k-1} - \ldots -a_n^{(k)}y_{k-n}+b_1^{(k)}u_{k-1}+\ldots \nonumber \\
&+b_m^{(k)}u_{k-m}+c_1^{(k)}e_{k-1} + \ldots + c_p^{(k)}e_{k-p} \nonumber \\
e_k =& y_k-\hat{y}_k \label{eq:217}
\end{align}
for $k\ge0$, with $y_k=0, u_k=0, e_k = 0$ for all $k<0$. Following Lemma~\ref{lem:0}, its state-space realisation is given by
\begin{align}
\hat{x}_{k+1} &=A^{(k)}\hat{x}_k+ B_1^{(k)}u_k+B_2^{(k)}e_k \nonumber \\
&=\hspace{-1mm} \left [ \begin{array}{cccc}
0 & \ldots & 0 & -a_n^{(k)} \\ 1 & \ddots & \vdots & -a_{n-1}^{(k)}\\
\ & \ddots & 0 & \vdots \\ 0 & \ldots & 1 & -a_1^{(k)}\end{array}\hspace{-1mm}\right ]\hspace{-1mm} \hat{x}_k + \hspace{-1mm} \left [\begin{array}{c} 0 \\ b_m^{(k)} \\ \vdots \\ b_1^{(k)} \end{array}\hspace{-1mm}\right ]\hspace{-1mm} u_k + \hspace{-1mm} \left [\begin{array}{c} \tilde{c}_n^{(k)} \\  \tilde{c}_{n-1}^{(k)} \\ \vdots \\ \tilde{c}_1^{(k)} \end{array}\hspace{-1mm}\right ]\hspace{-1mm} e_k\nonumber \\
y_k & = C\hat{x}_k+e_k = [0 \ \ldots \ 0 \ 1]\hat{x}_k + e_k. \label{eq:218}
\end{align}

\begin{thm}\label{thm:4}
Under the same conditions as in Theorem~\ref{thm:3}, the state-space realisation (\ref{eq:218}) approaches the optimal state estimator of (\ref{eq:3}) asymptotically. That is, defining the state estimation error $\varepsilon_k = x_k - \hat{x}_k$ between the states of (\ref{eq:3}) and (\ref{eq:218}), then the estimation error covariance  
\begin{align} 
\Sigma_k &= \mathbb{E}[\varepsilon_k\varepsilon_k^T]\rightarrow 0 \ \mathrm{as\ } k\rightarrow \infty. \label{eq:219}
\end{align}
\end{thm}

\begin{IEEEproof}
From (\ref{eq:3}) and (\ref{eq:218}), the estimation error dynamics is given by
\begin{align*}
\varepsilon_{k+1} = Ax_k -A^{(k)}\hat{x}_k +(B_1-B_1^{(k)})u_k + B_2w_k - B_2^{(k)}e_k.
\end{align*}
As $k\rightarrow \infty$, the above approaches
\begin{align*}
\varepsilon_{k+1} &\rightarrow  A\varepsilon_k +B_2(w_k -e_k) \\
&=A\varepsilon_k +B_2(w_k -y_k +C\hat{x}_k) \\
&=A\varepsilon_k -B_2C\varepsilon_k\\
&= (A-B_2C)\varepsilon_k. 
\end{align*}
From (\ref{eq:216}), the above is stable, hence $\Sigma_k\rightarrow 0$ as $k\rightarrow \infty$.
\end{IEEEproof}

\section{Model-free LQG Control for ARMAX Systems}

In this section, we apply the model-free state estimation results in the previous section to LQG control. We first give a result for model-based LQG control, then derive an algorithm for model-free LQG control. 

\subsection{Model-based LQG Control}

Consider the system model (\ref{eq:3}) and the value function
\begin{align}
V_{\pi}(x_k) &= \mathbb{E}[\sum_{t=k}\gamma^{t-k}(x_t^TQx_t+u_t^TRu_t)]\label{eq:301}
\end{align}
with discount factor $0<\gamma<1$, $Q\ge0$ and $R>0$. The objective is to design a stationary control policy $u_t=\pi(y_{<t})$ to minimise $V_{\pi}(x_k)$. This is a generalisation of the deterministic LQR problem studied in \cite{Lewis} where the noise $w_k$ void and the state $x_k$ is available. The discount factor is necessary to ensure the boundedness of the value function. 

We have the following result.

\begin{prop}\label{prop:3}
Consider the system (\ref{eq:3}) and the value function (\ref{eq:301}). 
The optimal control policy is given by  
\begin{align}
u_k &= u_k^{\star}= K\hat{x}_k, \label{eq:302}
\end{align}
where $\hat{x}_k$ is the optimal estimate of $x_k$ based on $y_{<k}$ and $K$ is given by 
\begin{align}
K &= (B_1PB_1^T+\gamma^{-1}R)^{-1}B_1^TPA \label{eq:303}
\end{align}
with $P$ being the solution to the discrete-time algebraic Riccati equation (DARE):
\begin{align}
P&=Q+\gamma \{A^TPA-A^TPB_1(B_1^TPB_1+\gamma^{-1}R)^{-1}B_1^TPA\},\label{eq:304}
\end{align}
and the optimal value function in steady state is given by 
\begin{align}
V_{\star}(x_k) &= x_k^TPx_k + \frac{\gamma\sigma^2}{1-\gamma}B_2^TPB_2. \label{eq:305}
\end{align}
\end{prop}
(See Appendix for proof.)

\subsection{Model-free LQG Control}

We now solve the model-free LQG control problem. 

For any control policy $\pi$, define the $Q$-function \cite{Lewis} as follows:
\begin{align}
Q_{\pi}(x_k,u_k)&= x_k^TQx_k+u_k^TRu_k+\gamma \mathbb{E}[V_{\pi}(x_{k+1})] \label{eq:306}
\end{align}
For the optimal policy $\pi^{\star}$, using the optimal value function in steady state (\ref{eq:305}), we get 
\begin{align*}
&Q_{\star}(x_k,u_k)\\
=&x_k^TQx_k+u_k^TRu_k+\gamma \mathbb{E}[V_{\star}(x_{k+1})] \\
=& x_k^TQx_k+u_k^TRu_k+\frac{\gamma^2\sigma^2}{1-\gamma}B_2^TPB_2 \\
&+\gamma \mathbb{E}[(Ax_k+B_1u_k+B_2w_k)^TP(Ax_k+B_1u_k+B_2w_k)]\\
=&x_k^TQx_k+u_k^TRu_k+\gamma (Ax_k+B_1u_k)^TP(Ax_k+B_1u_k)  \\
&+\frac{\gamma^2\sigma^2}{1-\gamma}B_2^TPB_2 +\gamma\sigma^2B_2^TPB_2 \\
=&[x_k^T\ u_k^T]\left [ \begin{array}{cc} H_{11}  & H_{12} \\
H_{12}^T & H_{22} \end{array}\right ] \left [ \begin{array}{c}
x_k \\ u_k \end{array}\right ] + \frac{\gamma \sigma^2}{1-\gamma} B_2^TPB_2.
\end{align*}
where $H_{11}=Q+\gamma A^TPA$, $H_{12}=\gamma A^TPB_1$ and $H_{22}=\gamma B_1^TPB_1+R$. 

Then, minimising $Q_{\star}(x_k,u_k)$ with respect to $u_k$ yields the optimal $u_k^{\star}$ and $Q_{\star}$ in steady state:
\begin{align}
u_k &= -H_{22}^{-1}H_{21} \mathbb{E}[x_k|y_{<k}] = K\hat{x}_k =u_k^{\star} \label{eq:307}\\
Q_{\star}(x_k,u_k^{\star}) &=x_k^T (H_{11}-H_{12}H_{22}^{-1}H_{12}^T)x_k + \frac{\gamma \sigma^2}{1-\gamma} B_2^TPB_2 \nonumber \\
&= x_k^TPx_k + \frac{\gamma \sigma^2}{1-\gamma} B_2^TPB_2 \label{eq:308}
\end{align}
by using (\ref{eq:303})-(\ref{eq:304}). 

Using the proposed on-line identification algorithm (Algorithm~\ref{alg:1}) and the model-free state estimation result (Theorem~\ref{thm:4}), we can use the following recursion for approximating $P$ and $K$:
\begin{align}
\hspace{-1mm}P_{k+1} =&\ Q+\gamma \{(A^{(k)})^TP_kA^{(k)}-(A^{(k)})^TP_kB_1^{(k)}\nonumber \\
&\cdot((B_1^{(k)})^TP_kB_1^{(k)}\hspace{-1mm}+\gamma^{-1}R)^{-1}(B_1^{(k)})^TP_kA^{(k)}\}, \label{eq:309}
\end{align}
with any $P_0>0$, and 
\begin{align}
K_k &= (B_1^{(k)}P(B_1^{(k)})^T+\gamma^{-1}R)^{-1}(B_1^{(k)})^TP_kA^{(k)}. \label{eq:310}
\end{align}

We have the following result on model-free LQG control.
\begin{thm}\label{thm:5}
Consider the system (\ref{eq:3}) and the value function (\ref{eq:301}). Let the model-free LQG control policy $\pi_k$ be 
\begin{align}
u_k &= K_k \hat{x}_k \label{eq:310}
\end{align}
with $K_k$ as in (\ref{eq:309})-(\ref{eq:310}) and $\hat{x}_k$ as in (\ref{eq:218}). Then, we have $\pi_k\rightarrow \pi^{\star}$ (the optimal policy), i.e., $K_k\rightarrow K$, $P_k\rightarrow P$, $u_k\rightarrow u_k^{\star}$, $V_{\pi_k}(x_k)\rightarrow V_{\star}(x_k)$ for all $x_k$, as $k\rightarrow \infty$. 
\end{thm}

\begin{IEEEproof}
We first claim that $P_k>0$ for all $k\ge0$ and $\lim_{k\rightarrow \infty}P_k=P$ if (\ref{eq:309}) is modified to 
\begin{align}
 P_{k+1}=Q+\gamma \{&A^TP_kA\nonumber \\
 &-A^TP_kB_1(B_1^TP_kB_1+\gamma^{-1}R)^{-1}B_1^TP_kA\}.\label{eq:311}
\end{align}
To see the claim above, we note that the above iteration can be rewritten (using the matrix inversion lemma) as
\begin{align*}
 P_{k+1}=Q+\gamma A^T(P_k^{-1}+\gamma B_1R^{-1}B_1^T)^{-1}A.
\end{align*}
This immediately leads to $P_k\ge0$ for all $k\ge0$. In fact, it is known \cite{Anderson} that, under the assumption that $(A, B_1)$ is controllable and $P_0>0$, $P_k>0$ for all $k\ge0$ and $\lim_{k\rightarrow \infty}P_k=P$. Note that $(A, B_1)$ is indeed controllable because (\ref{eq:3}) is a minimal realisation. Hence, the claimed property holds.

Next, from Proposition~\ref{prop:3}, we have $A^{(k)}\rightarrow A$, $B_1^{(k)}\rightarrow B_1$ and $B_2^{(k)}\rightarrow B_2$ as $k\rightarrow \infty$. It is easy to see that as $k\rightarrow \infty$, (\ref{eq:309}) converges to (\ref{eq:311}), hence $\lim_{k\rightarrow \infty}P_k=P$ still holds for (\ref{eq:309}), which further implies $K_k\rightarrow K$, hence, $\pi_k\rightarrow \pi^{\star}$. 
\end{IEEEproof}


\section{Conclusion}

Model-free state estimation is a challenging problem due to the fact that both the system model is unknown and the measurement contains partial state and noises.  By reformulating the classical system identification as a reinforcement learning problem and incorporating the classical tools of instrumental variables and bootstrapping, we have provided a value-iteration based reinforcement learning algorithm for system identification of an ARMAX system with guaranteed consistency. This algorithm is then used in solving the model-free state estimation problem for an ARMAX system, and a reinforcement learning solution has been obtained. These results have also been applied to solving the model-free LQG problem for an ARMAX system. 

The key to our model-free state estimation solution is to use the observable-canonical realisation, which leads to the optimal state estimation by the driving the state estimation error covariance to zero. How to generalise this observation to more general systems, linear or nonlinear, will be crucial to more general solutions to model-free state estimation. This will alleviate a stumbling block to reinforcement learning applications where only measurement of partial state with noise is available. 

\section*{Appendix A: Proof of Lemma~\ref{lem:0}}

\begin{IEEEproof}
Extend $b_n=\ldots =b_{m+1}=0$. From (\ref{eq:3}), we get 
\begin{align}
zx_{k,1} +a_nx_{k,n} &= b_n u_k + \tilde{c}_n w_k\nonumber \\
-x_{k,1}+ zx_{k,2} +a_{n-1}x_{k,n} &= b_{n-1} u_k + \tilde{c}_{n-1} w_k\nonumber  \\
-x_{k,2}+ zx_{k,3} +a_{n-2}x_{k,n} &= b_{n-2} u_k + \tilde{c}_{n-2} w_k \nonumber \\
\ldots & \ldots \nonumber \\
-x_{k,n-1} +(z+a_1)x_{k,n} &= b_1 u_k + \tilde{c}_1 w_k \label{eq:3-1}
\end{align}
Multiplying the first row above by $z^{-1}$ and adding it the second row, we get
\begin{align*}
&zx_{k,2} +(a_{n-1}+a_nz^{-1})x_{k,n} \\
=& (b_{n-1} +b_nz^{-1})u_k + (\tilde{c}_{n-1} +\tilde{c}_nz^{-1})w_k
\end{align*}
Again, multiplying this row by $z^{-1}$ and adding it to the third row in (\ref{eq:3-1}), 
we get 
\begin{align*}
&zx_{k,3} +(a_{n-2}+a_{n-1}z^{-1}+a_nz^{-2})x_{k,n} \\
=& (b_{n-2}+b_{n-1}z^{-1} +b_nz^{-2})u_k \\
&+ (\tilde{c}_{n-2}+\tilde{c}_{n-1}z^{-1} +\tilde{c}_nz^{-2})w_k
\end{align*}
Repeating this until the final row of (\ref{eq:3-1}), we get
\begin{align*}
&(z+a_1+a_2z^{-1}+\ldots a_nz^{n-1})x_{k,n} \\
=& 
(b_1+b_2z^{-1}+\ldots b_nz^{n-1})u_k \\
&+ (\tilde{c}_1+\tilde{c}_2z^{-1} + \ldots + \tilde{c}_nz^{n-1})w_k 
\end{align*}
Multiplying the above by $z^{-1}$ again, we get
\begin{align*}
a(z) x_{k,n} &= b(z) u_k + \tilde{c}(z)w_k 
\end{align*}
where $\tilde{c}(z) = \tilde{c}_1z^{-1} +\ldots +\tilde{c}_nz^{-n}$. It follows from (\ref{eq:3}) that 
\begin{align*}
a(z) y_k & = a(z)x_{k,n} +a(z) w_k \\
&=  b(z) u_k + (a(z)+\tilde{c}(z))w_k \\
&= b(z) u_k + c(z)w_k.
\end{align*}
Hence (\ref{eq:3}) is a state-space realisation of (\ref{eq:2}). 
\end{IEEEproof}  

\section*{Appendix B: Proof of Proposition~\ref{prop:1}}
\begin{IEEEproof}
We prove by contradiction. Suppose $R_y$ is rank deficient. Then, there exists some vector $v=[v_1\ \ldots \ v_n]^T\ne0$ such that $R_yv=0$. We first consider the case the first element of $v$, $v_1\ne0$ and take $v_1=1$ without loss of generality. 

It is easy to verify that the $(i+1)$-th row of $R_y, i=0,1, \ldots, n-1$ is given by
\begin{align*}
R_{y,i} = [r_y(p+i) \ \ldots\  r_y(p+i-n+1)].
\end{align*} 
Then, $R_{y,i}v =0$ for all $i=0,1,\ldots, n-1$. 

For any $i\ge1$, it holds that
\begin{align*}
r_y(p+i) &= \mathbb{E}[y_{k-p-i}y_k] \\
&=\mathbb{E}[y_{k-p-i}(-a_1y_{k-1} - \ldots -a_ny_{k-n}\\
&\hspace{20mm}+w_k +c_1w_{k-1}+\ldots c_pw_{k-p})]\\
&=\mathbb{E}[y_{k-p-i}(-a_1y_{k-1} - \ldots -a_ny_{k-n})]\\
&=-[a_n \ \ldots a_1][r_y(p+i-n) \ \ldots \ r_y(p+i-1)]^T.
\end{align*}
It follows that 
\begin{align*}
[r_y(p+n)  \ldots r_y(p+1)] = -[a_n  \ \ldots a_1]R_y,
\end{align*}
giving $R_{y,n} = [r_y(p+n)  \ \ldots r_y(p+1)]v=0$. That is, we have extended $R_y$ by one row at the bottom and still maintains its rank deficiency. The above process can be repeated indefinitely to give the result that 
\begin{align}
R_{y,i}=[r_y(p+i)  \ \ldots r_y(p+i-n+1)]v=0, \ \forall i\ge0. \label{eq:A1}
\end{align}
Denoting $V(z) = v_1 + v_2z^{-1}+\ldots v_nz^{-(n-1)}$ and the one-sided $Z$-transform of $r_y(k)$ as 
\begin{align*}
\hat{r}_y(z) &= r_y(0) + r_y(1)z^{-1} + r_y(2)z^{-2} + \ldots 
\end{align*}
Then, using (\ref{eq:A1}) and $Z$-transform properties, we get 
\begin{align*}
V(z) \hat{r}_y(z) =  D(z) = d_0 + d_1z^{-1} + d_{p-1} z^{-p}
\end{align*}
where $d_0, \ldots d_{p-1}$ depend on $r_y(0), \ldots, r_y(p-1)$. It follows that the spectrum of $y_k$ is given by 
\begin{align*}
S_y(z) &= \sum_{k=-\infty}^{\infty} r_y(|k|)z^{-k} \\
&= \hat{r}_y(z) + \hat{r}_y(z^{-1}) - r_y(0) \\
&= \frac{D(z)}{V(z)} + \frac{D(z^{-1})}{V(z^{-1})} - r_y(0).
\end{align*}
But from (\ref{eq:1}) (without $u_k$), the spectrum should be given by
\begin{align*}
S_y(z) = \frac{c(z)c(z^{-1})}{a(z)a(z^{-1})}\sigma^2
\end{align*}
These two expressions have a clear mismatch of the order in the denominator because $a(z)$ is $n$-th order and $V(z)$ is $(n-1)$-th order). This contradiction implies that $R_y$ can not be rank deficient. 

A similar proof works if $v_1=\ldots v_{j-1}=0$ for $j>1$ but $v_j\ne0$. But the details are omitted.
\end{IEEEproof}

\section*{Appendix C: Proof of Proposition~\ref{prop:3}}

\begin{lem}\label{lem:C}
Consider the system 
\begin{align}
x_{k+1}&=Ax_k + Bw_k\label{A:3}
\end{align}
with stable $A$ and $w_k\sim \mathcal{N}(0,\sigma^2)$, and the value function
\begin{align}
V(x_k) &= \mathbb{E}[\sum_{t=k}^{\infty} \gamma^{t-k} x_t^TQx_t] \label{A:4}
\end{align}
with $Q\ge0$. Then, 
\begin{align}
V(x_k) &= x_k^TPx_k + \frac{\gamma\sigma^2}{1-\gamma}B^TPB \label{A:5}
\end{align}
with 
\begin{align}
P&= \sum_{k=0}^{\infty} \gamma^k (A^k)^TQA^k = Q+\gamma A^TPA. \label{A:6}
\end{align}
\end{lem}
\begin{IEEEproof}
It is straightforward to verify that 
\begin{align*}
&\ \ \ V(x_k) \\
&= x_k^TQx_k + \mathbb{E}[\sum_{t=k+1}^{\infty} \gamma^{t-k} x_t^TQx_t] \\
&= x_k^TQx_k +\gamma \mathbb{E}[\sum_{t=k}^{\infty} \gamma^{t-k}x_{t+1}^TQx_{t+1}] \\
&= x_k^TQx_k +\gamma \mathbb{E}[\sum_{t=k}^{\infty} \gamma^{t-k}(Ax_t+Bw_t)^TQ(Ax_t+Bw_t)] \\
&= x_k^TQx_k + \gamma \sum_{t=k}^{\infty} \gamma^{t-k}B^TQB\sigma^2 \\
&\ \ \ +\gamma \mathbb{E}[\sum_{t=k}^{\infty} \gamma^{t-k}x_t^T(A^TQA)x_t]\\
&= x_k^TQx_k +\frac{\gamma\sigma^2}{1-\gamma}B^TQB+\gamma \tilde{V}(x_k)
\end{align*}
where 
\begin{align*}
\tilde{V}(x_k)&=\mathbb{E}[\sum_{t=k}^{\infty} \gamma^{t-k}x_t^T\tilde{Q}x_t] 
\end{align*}
with $\tilde{Q}=A^TQA$. Do the above repeatedly, we get
\begin{align*}
V(x_k) =& x_k^TQx_k+\gamma x_k^TA^TQAx_k + \gamma^2 x_k^T(A^2)^TQA^2x_k+\ldots \\
&+\frac{\gamma\sigma^2}{1-\gamma}(B^TQB+\gamma B^TA^TQAB+\ldots )\\
&= x^TPx_k + \frac{\gamma\sigma^2}{1-\gamma} B^TPB 
\end{align*}
with $P$ given by the first part of (\ref{A:6}). The second part of (\ref{A:6}) is then easily verified and the convergence of the sum in (\ref{A:6}) is guaranteed by the stability of $A$.
\end{IEEEproof}

Now we are ready to prove Proposition~\ref{prop:3}. 

\begin{IEEEproof} We first consider the state feedback case where $x_k$ is available. This is an infinite-horizon linear quadratic control problem with Gaussian noise. It is well known \cite{Anderson2} that the optimal control policy $\pi_{\star}$ is given by $u_k=Kx_k$ for some stabilising $K$, i.e., $\tilde{A}=A+B_1K$ is stable. Invoking Lemma~\ref{lem:C}, the corresponding value function is given by 
\begin{align*}
V_{\star}(x_k) &= x_k^TPx_k + \frac{\sigma^2}{1-\gamma}B_2^TPB_2
\end{align*}
with 
\begin{align*}
P = Q+\gamma(\tilde{A}^TP\tilde{A}) = Q+\gamma((A+B_1K)^TP(A+B_1K)).
\end{align*}
It is clear that the value function is minimised when $P$ is minimised by $K$. It is a well-known in optimal control \cite{Anderson2} that the optimal $P$ is given by the DARE (\ref{eq:304}) with the optimal $K$ is given by (\ref{eq:303}). 

When the state is not available, the well-known separation principle holds \cite{Anderson2} which says that that the optimal control policy is given by (\ref{eq:302}) with $\hat{x}_k$ being the optimal state estimate. We see from Theorem~\ref{thm:3} that the optimal state estimate $\hat{x}_k$ has zero estimation error covariance in steady state. Hence, (\ref{eq:305}) holds. 
\end{IEEEproof}


\begin{thebibliography}{10}
\bibitem{Ljung} L. Ljung. System Identification: Theory for the User, Prentice Hall, 2nd Edition,1999.
\bibitem{Soderstrom} T. S\"oderstr\"om and P. G. Stoica, Instrumental Variable Methods for System Identification, Springer-Verlag, 1983.
\bibitem{Anderson} B. D. O. Anderson and J. Moore, Optimal Filtering, Prentice Hall, 1979. 
\bibitem{Goodwin} G. C. Goodwin and K. S. Sin, Adaptive filtering prediction and control, Prentice Hall, 1984.
\bibitem{Silver} D. Sliver, Introduction of Reinforcement Learning with David Silver, Lecture Series, DeepMind 2015. (https://deepmind.com/learning-resources/-introduction-reinforcement-learning-david-silver)
\bibitem{Bertsekas} D. Bertsekas, Reinforcement Learning and Optimal Control, Athena Scientific, 2019.
\bibitem{Lewis} F. Lewis and D. Vrabie, ``Reinforcement learning and adaptive dynamic programming for feedback control," IEEE Circuits and Systems Magazine, 9(3):32-50,  2009.
\bibitem{Durbin} Efficient estimators of parameters in moving-average models, {\em Biometrica}, 46:306-316, 1959.
\bibitem{Walker} A. M. Walker, Large-sample estimation of parameters for moving-average models,  {\em Biometrica}, 48:343-357, 1961.
\bibitem{Anderson2} B. D. O. Anderson and J. Moore, Optimal Control: Linear Quadratic Methods, Prentice Hall 1971.
\end{thebibliography}
\end{document}